\documentclass[9pt,twocolumn,twoside]{osajnl}

\journal{jocn} 

\usepackage{amsmath,amssymb,amsfonts}
\usepackage{graphicx}
\usepackage{textcomp}
\usepackage{xcolor}
\usepackage{lipsum}
\usepackage{subfig}
\setboolean{shortarticle}{false}

\title{Transformer-based Nonlinear Equalization for DP-16QAM Coherent Optical Communication Systems}

\author[1,2,*]{Naveenta Gautam}
\author[2,3]{Sai Vikranth Pendem}
\author[1,3]{Brejesh Lall}
\author[1,2,3]{Amol Choudhary}

\affil[1]{Bharti School of Telecommunication Technology and Management, Indian Institute of Technology Delhi, India}
\affil[2]{UltraFast Optical Communications and High Performance Integrated Photonics (UFO-CHIP) Group, Indian Institute of Technology Delhi, India}
\affil[3]{Department of Electrical Engineering, Indian Institute of Technology Delhi, India}

\affil[*]{Naveenta.Gautam@dbst.iitd.ac.in}

\begin{abstract}
Compensating for nonlinear effects using digital signal processing (DSP) is complex and computationally expensive in long-haul optical communication systems due to intractable interactions between Kerr nonlinearity, chromatic dispersion (CD), and amplified spontaneous emission (ASE) noise from inline amplifiers. The application of machine learning architectures has demonstrated promising advancements in enhancing transmission performance through the mitigation of fiber nonlinear effects. In this paper, we apply a Transformer-based model to dual-polarisation (DP)-16QAM coherent optical communication systems. We test the performance of the proposed model for different values of fiber lengths and launched optical powers and show improved performance compared to the state-of-the-art digital backpropagation (DBP) algorithm, fully connected neural network (FCNN) and bidirectional long short term memory (BiLSTM) architecture.

\end{abstract}

\setboolean{displaycopyright}{false} 

\begin{document}

\maketitle

\section{Introduction}
Coherent communication systems play a crucial role in enabling the implementation of advanced modulation schemes, which require higher average optical power per symbol \cite{kikuchiDSP}. However, this increased power makes the system more susceptible to nonlinear effects. In long haul coherent optical communication systems, the presence of Kerr nonlinear effects has been a major obstacle in achieving higher data rates \cite{NLOGPA}. While digital signal processing (DSP) techniques can effectively characterize and compensate for deterministic nonlinear effects, addressing the stochastic effects resulting from the interplay between Kerr nonlinearity, chromatic dispersion (CD), and amplified spontaneous emission (ASE) noise from inline amplifiers is challenging. To meet the growing demand for higher data rates, it is essential to find ways to mitigate the impact of nonlinear effects and compensate for their influence \cite{MLforNLC1}.

The most promising technique for nonlinear compensation in single channel systems is the digital backpropagation (DBP) algorithm, which backpropagates the received signal using inverted fiber parameters after numerically simulating the fiber channel. However, it has some disadvantages; firstly, it exhibits high computational complexity, making it less suitable for real-time applications. Additionally, it struggles to precisely represent the channel due to the presence of random parameters. As a result, the existing techniques used for nonlinearity compensation have proven to be complex and demanding in terms of computational resources, particularly in long haul links \cite{theoreticalML}. Consequently, the research focus has shifted towards machine learning (ML)-based approaches for nonlinear compensation, which offer promising alternatives \cite{MLforNLC1,MLinOC,fieldlab,CNN2,optidistillnet,gautam2021comparative}.

\begin{figure*}
\centering
\includegraphics[width=0.75\textwidth]{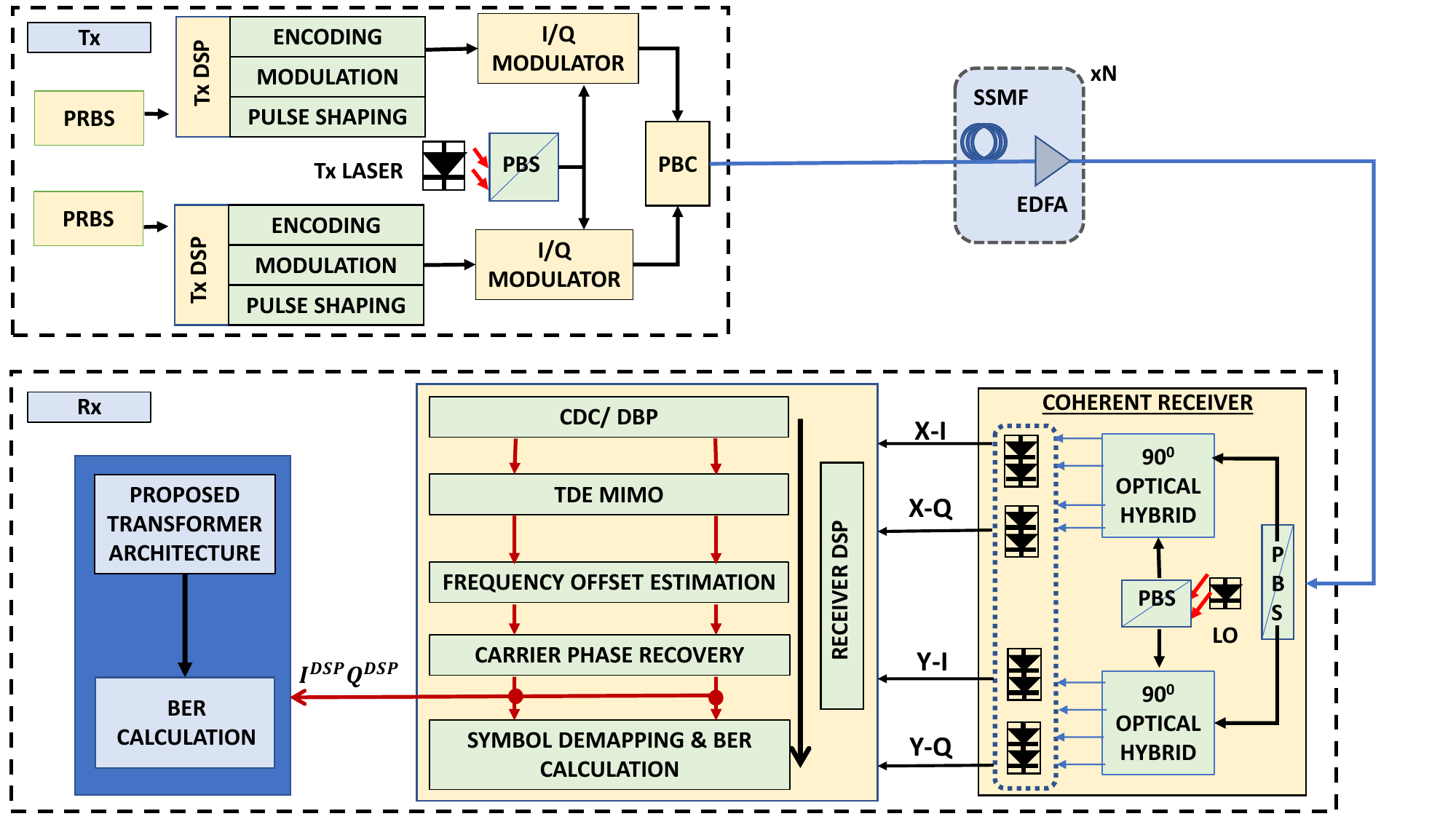} 
\caption{System Setup of a dual polarisation (DP)-16QAM coherent optical communication system.}
\label{sys setup}
\end{figure*}

Numerous machine learning (ML) techniques have been explored to address the challenge of nonlinear compensation (NLC) in optical fibers. These include neural networks (NNs) \cite{theoreticalML, fieldlab, MLforNLC1,eliasANN}, support vector machines (SVMs) \cite{eliasSVM, eliasunsupervisedSVM}, expectation maximization (EM) \cite{zibardispersionmanaged}, clustering techniques, and message passing algorithms \cite{futureinternet}. Although investigations into deep neural networks trained with symbol triplets \cite{fieldlab} were conducted for NLC, these approaches could not outperform the digital backpropagation (DBP) technique \cite{NNsea}. It's noteworthy that employing symbol triplets as input increased the input dimensionality, consequently elevating the overall complexity. Additionally, attempts were made to employ unsupervised SVMs for nonlinear equalization in both single and wavelength division multiplexing (WDM) scenarios; however, this technique could not outperform DBP in the context of single-channel systems \cite{eliasunsupervisedSVM}. Authors in \cite{DBPDNN, DLNLSE}, treat the linear and nonlinear stages of DBP as an integrated deep neural network, resulting in a marked reduction in complexity. Nevertheless, these methods were predominantly applied to dispersion managed or neglected links \cite{eliasANN}, or they omitted the consideration of various transmission impairments like polarization effects, WDM effects, laser-phase noise, and laser-frequency offsets \cite{eliasANN,wang2015nonlinear, eliasSVM,zibardispersionmanaged}.


A neural network (NN) architecture that has been recently unveiled by Google AI and is gaining popularity is the Transformer \cite{attentionisallyouneed}. One of the main reasons for its rising popularity is its self-attention mechanism, which aids models in focusing on certain aspects of input and reason more effectively. It captures the global dependencies in the input and output. Transformer uses an encoder and decoder architecture but eliminates recursion in favour of attention mechanisms, which enables substantially greater parallelization than other networks such as recurrent and convolutional NNs \cite{attentionisallyouneed}. Inspired by the great success of Transformer for machine translation tasks in natural language processing (NLP), we apply the self-attention mechanism to learn the inverse mapping of the fiber channel by considering the received symbols to be sequential due to inter symbol interference noise in neighbouring symbols. A simplified version of Transformer has been combined with the feature decoupled distributed (FDD) scheme and introduced in optical orthogonal frequency division multiplexing (OFDM) systems for fast and accurate channel modeling \cite{transformerOFDM}. However, the authors do not utilise self-attention, which is the highlight of Transformer architecture. Transformer has also been used by authors in \cite{transNLE} for nonlinear channel equalization using dispersion precompensation at the transmitter but it does not outperform the DBP algorithm and lacks a comparison of the proposed model with existing NN architectures.

In this study, Transformer is used in coherent optical communication systems to perform nonlinear compensation to undo the effects of CD, nonlinearity, and ASE noise of the channel. Since the Transformer has difficulty in learning the mapping when the received symbols are directly fed, we use a float to bit conversion to increase the dimension of feature embeddings and capture relationships that might not be apparent from the raw data alone. Fiber nonlinearity increases in direct proportion to the launched optical power and transmission distance, therefor, we test the performance of Transformer by varying both these parameters. The modeling accuracy is represented in terms of Q factor gain with respect to (wrt) linear equalization, DBP, fully connected NN (FCNN) and bidirectional long short term memory (BiLSTM). Results demonstrate that Transformer outperforms FCNN and BiLSTM and can provide a gain of approximately 2 dB wrt linear equalisation and 1.65 dB wrt DBP. In addition, our solution has improved the signal quality for a received signal of Q factor as low as 7.2 dB. The proposed Transformer model also shows performance improvement when compared against LSTM. Although the training process is often time-consuming, testing is real-time since the transmitted symbols are computed by simply feeding the received signals to the trained network.

In this work, we address the nonlinear compensation problem for dual polarisation (DP)-16QAM coherent optical communication system for different fiber lengths and launched optical powers. The main contributions are:
\begin{itemize}
\item We integrate the Transformer architecture with the receiver side DSP algorithms in coherent optical communication systems. We propose a simple float to bit conversion method to increase the feature dimension and aid the model to learn the nonlinear mapping. 
\item We investigate the performance of Transformer for different input sequence lengths. 
\item The Transformer architecture is compared with FCNN and BiLSTM for varying fiber lengths and launched optical powers.
\end{itemize}

The rest of the paper is structured as follows. In Section~\ref{section2} we describe the system setup, proposed Transformer architecture, training data preparation and methodology. The results are presented and discussed in Section~\ref{sectionresults}, while Section~\ref{sectionconclusion} concludes the paper. 

\section{System Architecture and Methodology}  \label{section2}
In this section we describe the system architecture used for demonstrating our results. The proposed approach can be used for different system setups as well. Our work aims to integrate the Transformer architecture for nonlinear compensation at the receiver side of a optical coherent communication system. 

\subsection{System Setup} \label{systemsetup}

We model a DP-16QAM coherent optical communication system to prove the feasibility of the proposed Transformer based approach of improving the system performance. Figure. \ref{sys setup} shows the system setup and the receiver side DSP of a DP-16QAM coherent optical communication system. At the transmitter, a psuedo random bit sequence (PRBS) generator is used to generate data sequences for the two polarisations which are fed to the bit to symbol mapping block. Then it is upsampled to 8 samples per symbol for Nyquist pulse shaping with a roll off factor of $0.18$. This sequence is subsequently modulated using a carrier at 1550 nm with a linewidth of 100 KHz. The modulated dual polarisation signal is multiplexed and transmitted on a link consisting of multiple fiber spans of 80 km each and an erbium-doped fiber amplifier (EDFA) to compensate for the path loss. At the receiver, the optical signal is detected by a coherent receiver, and digital signal processing is done to compensate for chromatic dispersion, polarisation mode dispersion, frequency offset, carrier phase recovery, and constellation alignment. This is followed by thresholding and symbol demapping. The system parameters are summarized in Table~\ref{Sys parameters}

Signal propagation inside an optical fiber is governed by the nonlinear Schrödinger equation (NLSE) \cite{NLOGPA} which is given by:
\begin{equation}
  i\frac{\partial \psi}{\partial z}+ i\frac{\alpha}{2}\psi+ \frac{\beta_{2}}{2}\frac{\partial^2 \psi}{\partial t^2}-\frac{\beta_{3}}{6}\frac{\partial^3 \psi}{\partial t^3}+\gamma |\psi|^2 \psi = 0
  \label{eq1}
\end{equation}
where $\psi$ is the complex envelope of the optical field, $z$ is the distance, $\alpha$ is the attenuation, $t$ is the time coordinate, $\beta_{2}$ is the group velocity dispersion (GVD), $\beta_{3}$ is third order dispersion and $\gamma$ is the nonlinear parameter. The nonlinear parameter governs three main non linear effects namely self phase modulation, cross phase modulation and four wave mixing. 

\begin{table}[]
\centering
\caption{System Parameters}
\label{Sys parameters}
\begin{tabular}{|c|cl|}
\hline
\textbf{Network Parameters} & \multicolumn{2}{c|}{\textbf{Value}}    \\ \hline
Wavelength                  & \multicolumn{2}{c|}{1550nm}            \\ \hline
Symbol Rate                 & \multicolumn{2}{c|}{10 GBaud}          \\ \hline
Launched Optical Power      & \multicolumn{2}{c|}{-4 to 3 dBm}                  \\ \hline
Attenuation                 & \multicolumn{2}{c|}{0.2 dB/km}         \\ \hline
Dispersion                  & \multicolumn{2}{c|}{17 ps/(nm.km)}     \\ \hline
Core Area                   & \multicolumn{2}{c|}{80 $um^{2}$}       \\ \hline
Nonlinear Refractive index  & \multicolumn{2}{c|}{2.6e-8 $um^{2}/W$} \\ \hline
Span Length                 & \multicolumn{2}{c|}{80 km}             \\ \hline
Noise figure of EDFA        & \multicolumn{2}{c|}{4.5 dB}            \\ \hline
\end{tabular}
\end{table}

The variation of signal intensity during fiber propagation causes variations in the refractive index, which causes the signal phase to change. As a result, the nonlinear phase variation is self-induced, and the associated phenomenon is known as SPM \cite{NLOGPA}. This results in a frequency shift known as frequency chirping, which interacts with the dispersion in the optical fiber and broadens the optical pulse's spectrum. In transmission systems with a high input power, pulse broadening rises because the chirping effect is proportional to the launched optical power. 

\subsection{Transformer Architecture}

The Transformer is a type of NN that utilizes self-attention mechanism to weigh the importance of different parts of the input sequence. Self-attention can be understood as a mechanism for computing a weighted representation of each element in a sequence by considering the relationships between all elements in the sequence. This allows the Transformer to capture long-range dependencies and patterns in the input signal, making it a powerful tool for signal processing tasks such as audio and image processing, speech recognition, and time-series prediction \cite{attentionisallyouneed}. In recent years, several Transformer-based models have been developed specifically for signal processing applications, including the TransUnet \cite{transunet}, and the Waveformer \cite{waveformer}. These models have shown impressive results in various signal processing tasks, surpassing traditional methods and achieving state-of-the-art performance. 
 
The Transformer architecture consists of an encoder and decoder. The encoder's primary role is to process and extract meaningful features from the input data while the decoder's main purpose is to generate output sequences based on the learned feature representations from the encoder. The multi-head attention and feed forward network comprise the encoder part of the architecture. Multi-head self-attention layer computes self-attention among the input tokens, enabling each token to attend to other relevant tokens in the sequence. It captures dependencies and relationships between symbols, allowing the model to consider the global context while encoding information. After the self-attention sub-layer, a position-wise feed-forward network is applied to each token independently. It consists of two linear transformations with a nonlinear activation function in between. This operation helps to model more complex interactions between tokens and introduces nonlinearity into the network. Both the self-attention and feed-forward sub-layers have residual connections around them \cite{attentionisallyouneed} which allow the model to retain information from previous layers, mitigating the issue of vanishing gradients. Layer normalization is applied to stabilize the training process by normalizing the inputs of each sub-layer. To capture hierarchical features and create deeper representations, multiple Transformer encoder layers are stacked on top of each other. Each layer receives the output of the previous layer as its input. The Transformer decoder utilizes a self-attention mechanism, which allows it to focus on different parts of the input sequence to capture dependencies and relationships between sequences. During training, masked self-attention is applied to prevent the decoder from attending to future positions in the output sequence, ensuring autoregressive generation. Multiple attention heads are employed to capture diverse types of information, and positional encoding is used to consider sequence order and sequential information. Together, these elements enable the Transformer decoder to generate high-quality outputs for tasks like machine translation, language generation, and text summarization \cite{attentionisallyouneed}.

\begin{figure}[t]
	\centerline{\includegraphics[width=0.65\textwidth,height = 7.5cm]{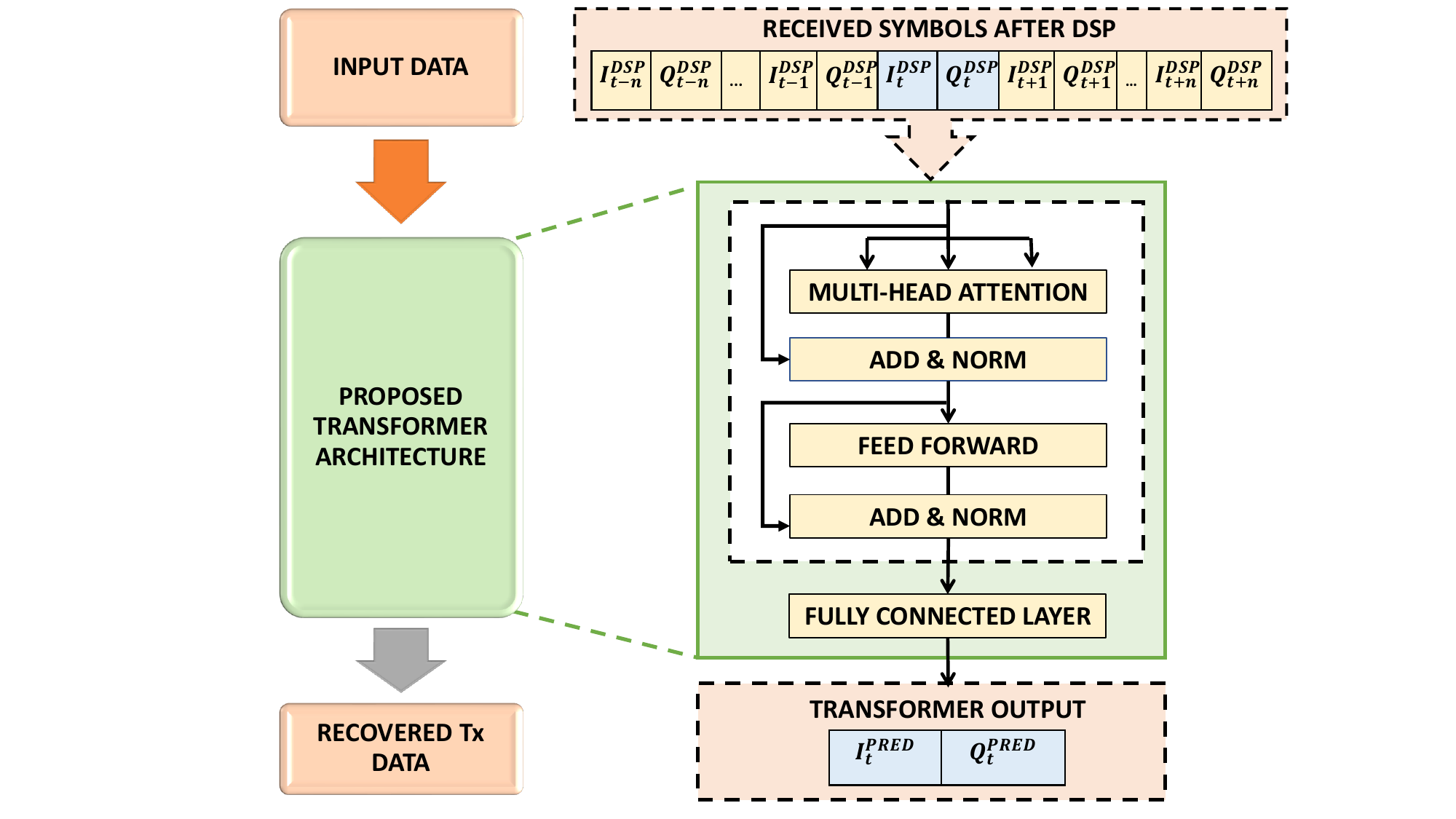}}
	\caption{Transformer architecture and training data preparation }
\label{trans architecture}
\vspace{-3ex}
\end{figure}

In certain signal processing tasks, like nonlinear equalization, the specific role of synthesis performed by the decoder in a Transformer architecture may not be directly applicable. These tasks often involve transforming and enhancing input signals rather than generating new sequences. Therefore, in cases where the generation of output sequences is not the primary objective, using only the encoder can be a valid approach. We simplify the Transformer and only use the encoder part followed by a fully-connected output layer to simplify the architecture and save computational resources as shown in Fig.~\ref{trans architecture}. We also remove the embedding layer since it was designed for word vectors and is unsuitable for this application. The purpose of the fully connected layer is to map the high-level representations obtained from the Transformer encoder layer to the desired output format, depending on the specific task at hand. Overall, the proposed architecture combines the power of the Transformer encoder layer, which captures contextual relationships in the input sequence, with the flexibility and mapping capabilities of the fully connected layer, enabling the model to learn complex patterns and make predictions for various tasks.

The self attention mechanism is implemented by capturing the relationships between the different elements (in this case, the past and future symbols) of the received sequence. The self-attention mechanism enables the inputs to interact with each other and determine who they should pay more attention to. These interactions and attention scores are aggregated as outputs.


\subsection{Training Data Preparation}

Tools based on ML are especially effective for handling equalization and nonlinear transformations and have been successfully used for nonlinear compensation \cite{optidistillnet,MLforNLC1,futureinternet}. We make an attempt to use Transformer for nonlinear equalization by accurately recovering transmitted symbols by observing the distorted symbols. The optical signals received by the coherent receiver contain information about the signal's optical path and any linear and nonlinear impairments it encounters along the path. 


Contrary to previous approaches where two separate NNs are used for recovering the real and imaginary part of the symbol \cite{futureinternet}, we use the in-phase (I) and quadrature (Q) components as a 2D vector for training. The I and Q components of the symbols were extracted after the DSP stage for training the Transformer model. To further capture the temporal correlation between adjacent symbols due to inter symbol interference generated by chromatic dispersion, we include $n$ symbols each from the past and future time steps. The sequence length of the training data is $2(2n+1)$. Initially, this sequence was directly input to the model. However, it became evident that the model encountered challenges in learning the input-to-output relationship due to limited feature dimensionality. To address this, the symbols within this input sequence were subsequently transformed from float64 to binary32, thereby expanding the size of the input features from 1 to 32. This conversion aimed to prevent underfitting caused by the constrained dimensionality. By providing additional features, Transformer learns more relevant and meaningful representations. The target vector consists of the I and Q values of the present symbol as shown in Fig.~\ref{trans architecture}.
\begin{equation}
   [I_{t-n}, Q_{t-n} \ldots, I_{t}, Q_{t}, \ldots I_{t+n}, Q_{t+n}]
\end{equation}

The optimal window length $n$ can vary depending on the specific characteristics of the optical fiber system and the nature of the nonlinear equalization task. We experiment with different window sizes and the results are presented in the next section.

\section{Results and Discussion}\label{sectionresults}




VPItransmissionMaker™ Optical Systems is used to simulate a single channel 10 GBaud dual polarisation (DP)-16QAM setup as described in section \ref{systemsetup}.
We use a total of $2^{19}$ symbols for training and generate the test data of the same size using a different random seed at the PRBS generator.
In this study, we aim to compare the performance of the Transformer architecture in terms of its impact on the improvement of the Q factor and the potential enhancement in fiber length for coherent optical communication systems. To provide a comprehensive analysis, we compare the Transformer with several existing methods, including linear equalization, DBP, FCNN and BiLSTM, which is another widely used time series architecture \cite{Ianbook}. For fair comparison, the same training dataset has been used for all the techniques.

The input tensors are provided in the form of (batch, sequence, features) for training the Transformer \cite{attentionisallyouneed}. For training we use a mini batch size of 1024, Adam optimiser with a learning rate of $10^{-3}$ and Rectified Linear Unit (ReLU) as the activation function. Transformer uses 8 heads, 1 encoder layer and 0.1 dropout. The parameters are summarized in Table \ref{Transformer table}. We have chosen mean square error (MSE) given by Eq. \ref{eqmse} as the loss function to update the weights.
    \begin{equation}\label{eqmse}
        MSE = \frac{1}{N} \sum_{i=1}^{N} (y_{i}-a_{i})^{2},
    \end{equation}
where $y$ is the desired output, $a$ is the calculated output and $N$ is the total number of outputs.

\begin{figure}[t]
	\centerline{\includegraphics[width=0.45\textwidth]{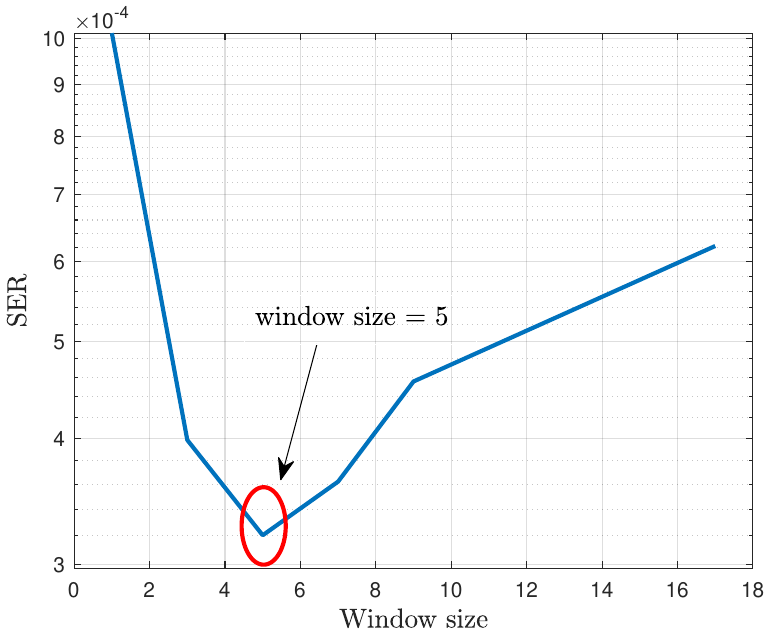}}
	\caption{SER vs window showing the optimum window length for our experiments}
\label{window length}
\vspace{-3ex}
\end{figure}

\begin{figure}[t]
	\centerline{\includegraphics[width=0.45\textwidth]{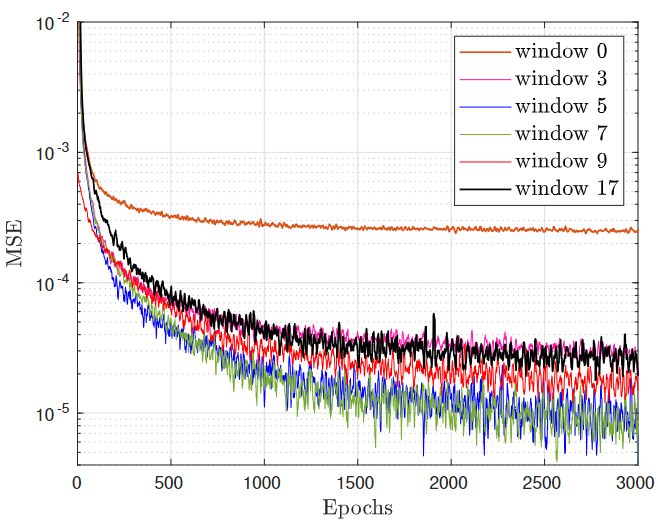}}
	\caption{Training curves for different window lengths}
\label{training curves for seq}
\vspace{-3ex}
\end{figure}

\begin{table}[t]
\centering
\caption{Transformer Architecture}
\label{Transformer table}
\begin{tabular}{|l|l|}
\hline
\textbf{Parameters}   & \textbf{Value} \\ \hline
Sequence Length       & 10             \\ \hline
Input features        & 32             \\ \hline
No. of heads          & 8              \\ \hline
No. of layers         & 1              \\ \hline
Feedforward dimension & 1024           \\ \hline
Dropout               & 0.1            \\ \hline
Output dimension      & 2             \\ \hline
\end{tabular}
\end{table}


\begin{table}[]
\centering
\caption{BiLSTM architecture}
\label{BiLSTM table}
\begin{tabular}{|c|cl|}
\hline
\textbf{Parameters}       & \multicolumn{2}{c|}{\textbf{Value}} \\ \hline
Input Dimension           & \multicolumn{2}{c|}{10}             \\ \hline
Number of LSTM layers     & \multicolumn{2}{c|}{2}              \\ \hline
Number of memory elements & \multicolumn{2}{c|}{4}              \\ \hline
Output dimension          & \multicolumn{2}{c|}{2}              \\ \hline
\end{tabular}
\end{table}

\begin{figure}[t]
	\centerline{\includegraphics[width=0.45\textwidth]{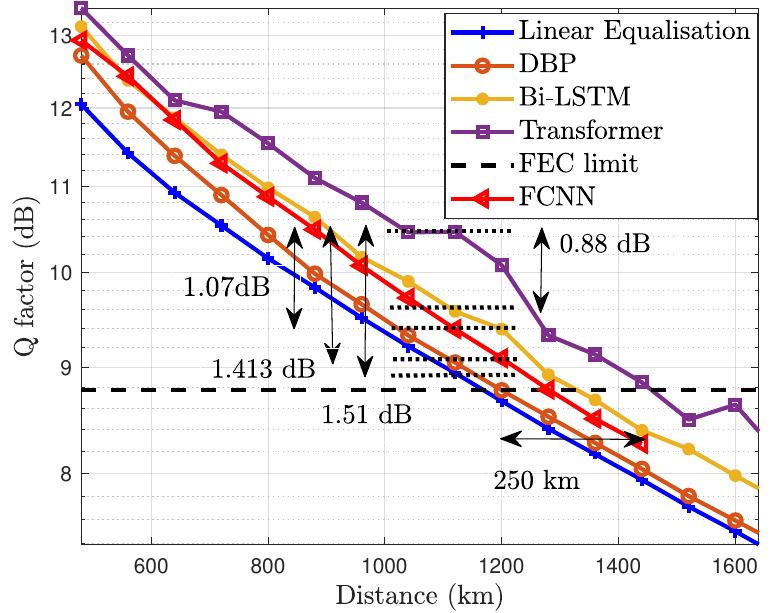}}
	\caption{Q factor Vs. Fiber length for DP-16QAM single channel system}
\label{length sweep}
\vspace{-3ex}
\end{figure}


As shown in Fig.~\ref{window length}, we experimented with window sizes ranging from 0 (no memory) to 17. Including a sufficient number of past and future symbols in the sequence allows the Transformer model to capture the contextual information necessary for accurate equalization. By incorporating past symbols, the model can account for the history of the signal, which may contain valuable information about previous interactions with the fiber medium. Including future symbols can enable the model to anticipate upcoming changes or distortions in the signal. Longer window lengths require more memory and computational resources to process. Transformer typically have a quadratic time and space complexity with respect to the sequence length. Therefore, increasing the sequence length significantly impacts the model's memory requirements and computational efficiency. It is important to strike a balance between the window length and available resources to ensure the model can be trained and deployed effectively. The window length in the context of using a Transformer for nonlinear equalization in optical fiber affects the model's ability to capture relevant context, its computational requirements, the balance between local and global information, and the risk of overfitting. While a larger sequence length can provide more information for the model to learn from, it also introduces challenges related to noise, overfitting, and model complexity that can lead to higher MSE if not properly managed. We chose a window size of 5 (2 past and 2 future symbols) for our experiments since it achieves the lowest SER, faster convergence during training and lower MSE as shown in Fig.~\ref{training curves for seq}. 

\begin{figure}[t]
	\centerline{\includegraphics[width=0.45\textwidth]{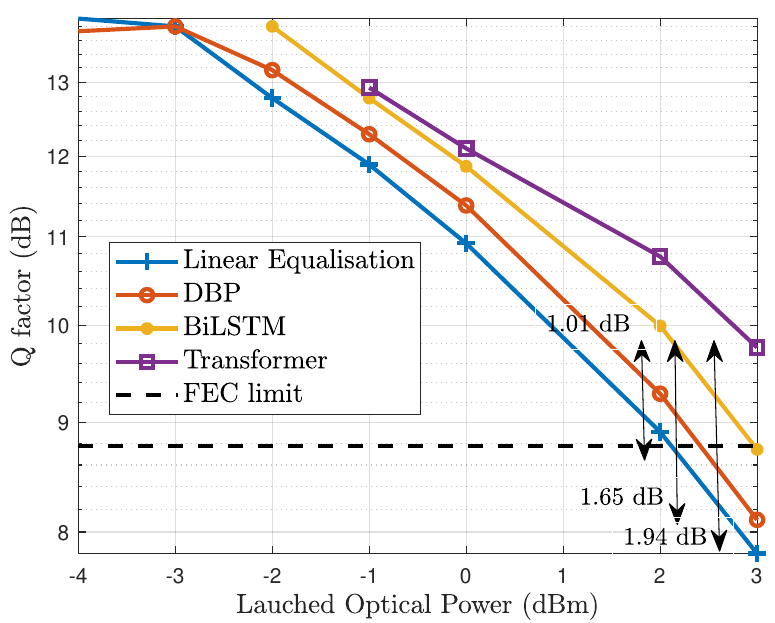}}
	\caption{Q factor Vs. LOP for DP-16QAM single channel system. BiLSTM and Transformer give 0 symbol errors corresponding to a BER of 0 for LOP in the range of -4 to -1 dBm}
\label{power sweep}
\vspace{-3ex}
\end{figure}

The coherently detected signal after DSP is fed to the trained models for further improvement in the BER. Figure. \ref{length sweep} and \ref{power sweep} show the measured Q-factor versus fiber length and launched optical power curve where the power and length are fixed at 0 dBm and 640 km respectively (system parameters specified in Table. \ref{Sys parameters}). The Q-factor is calculated from BER using Eq.~\ref{eq:1} \cite{ahmad2016radial}. 
\begin{equation}
\label{eq:1}
Q =20log_{10}[\sqrt2  erfc^{-1}(2BER)]
\end{equation}

We evaluated these methods on a simulated optical communication system with different levels of channel nonlinearity. Our results show that FCNN (single hidden layer with 100 neurons), BiLSTM (parameters summarized in Table \ref{BiLSTM table}) and Transformer outperform DBP (10 steps per span) in terms of BER performance across all tested fiber lengths and launched optical powers. However, Transformer achieves the best BER performance among all three methods. Specifically, at a fiber length of 1120 km, we achieve a maximum Q factor gain of 1.51dB, 1.41 dB, 1.07 dB and 0.88 dB with respect to linear equalisation, DBP, FCNN and LSTM, respectively as shown in Fig.~\ref{length sweep}. At a power of 3 dBm, when comparing the performance of the Transformer model with linear equalization, DBP, and BiLSTM the graph in Fig.~\ref{power sweep} shows a Q-factor gain of 1.94 dB, 1.65 dB and 1.01 dB, respectively. As the number of steps per span for DBP increases, the compensation of nonlinearities becomes more accurate at the expense of increased computational complexity \cite{NLOGPA}.  While the Transformer model suggested in \cite{transNLE} could not outperform DBP with 3 steps per span, our proposed technique outperforms DBP with 10 steps per span. 
Additionally, we found that Transformer provides a significant gain in the link length compared to both LSTM and DBP when operating below the forward error correction (FEC) limit. Transformer provides approximately 250 km link length gain compared to DBP and  150 km gain compared to BiLSTM, indicating that the Transformer-based model is not only more accurate but also more robust to noise and nonlinearity. Overall these Q-factor gains highlight the significant advantages of using the Transformer model over linear equalization and DBP in terms of improving the signal quality and reducing the impact of noise and distortion in the optical communication system.


The Transformer-based approach leverages its ability to capture complex patterns and dependencies in the signal, making it well-suited for modeling and mitigating fiber impairments. The FCNN and BiLSTM techniques follow closely behind the Transformer, suggesting its capability to learn and model the nonlinear dynamics of the optical channel. Although it may not outperform the Transformer, it still provides substantial improvement in terms of Q factor gains as compared to DBP and CDC. This indicates that while DBP can partially compensate for fiber impairments, it may not be as effective as the other techniques being evaluated. We can infer that linear equalisation alone may not be sufficient to achieve optimal performance in mitigating the effects of nonlinear phase noise on signal quality. 

\begin{figure}[t] 
    \centering
  \subfloat[After Bi-LSTM\label{1a}]{%
       \includegraphics[width=0.5\linewidth]{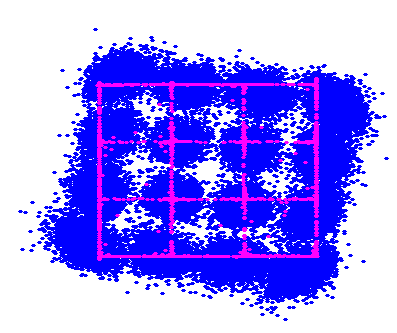}}
    \hfill
  \subfloat[After Transformer\label{1b}]{%
        \includegraphics[width=0.5\linewidth]{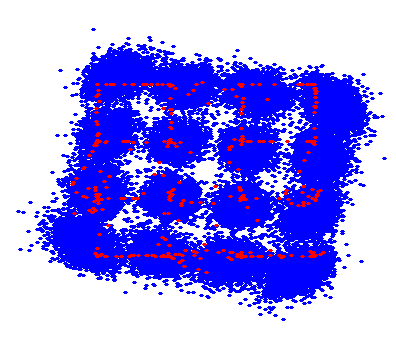}}
    \\
  \caption{Constellations before (blue) and after NLE using Bi-LSTM and Transformer for a LOP of 0dBm and 1200 km fiber length.}
  \label{constellation} 
\end{figure}
A NN based regressor tries to map the distorted signals to the ideal constellation points which leads to a squeezing effect in the constellation diagram after the Transformer and LSTM nonlinear equalizer as shown in Fig.~\ref{constellation} \cite{futureinternet}. The goal of this process is to map distorted symbols back towards the original ideal constellation points in the signal space to counteract the effects of nonlinear impairments. Therefor, a NN regressor tries to minimize the effects of nonlinear impairments by bringing the distorted signals closer to the ideal constellation points. This results in the distances between the estimated constellation points becoming smaller than their original separations, effectively compressing the constellation diagram. The superior performance of Transformer as compared with BiLSTM can also be inferred from the constellation diagram at a fiber length of 1200 km in Fig. \ref{constellation}. By training on data with nonlinear distortions, the model can learn to identify and correct these distortions in new, unseen data. This correction process helps to improve the accuracy and reliability of the signal, leading to better decision boundaries.

These findings suggest that advanced ML-based approaches such as Transformer, FCNNs and BiLSTMs hold promise in improving signal quality over long fiber lengths, surpassing traditional methods such as DBP and CDC. The improved performance for longer fiber lengths demonstrates the advantages of applying self-attention to understand and account for the temporal dynamics of the data. The Transformer learns to recover transmitted signals by adopting the self-attention mechanism and differentially weighting the significance of each symbol in the sequence of the input data. It is to be noted that for the DBP-based receiver, complexity increases with transmission distance, but we use the same Transformer architecture across all fiber lengths. This approach may have certain advantages, such as simplifying the implementation and reducing the need for adjustments based on transmission distance.

Transformer architectures are based on attention mechanism, which allows the network to focus on relevant parts of the input sequence. This results in more accurate predictions and better performance on long sequences of data. However, the self-attention mechanism used in Transformer can require more computations and parameters compared to LSTMs, resulting in a higher computational complexity. In practice, the computational complexity of both architectures can be reduced by optimizing the model architecture and using techniques such as pruning, quantization, and parallelization \cite{Ianbook,optidistillnet}. Nonetheless, the relative computational complexity of LSTM and Transformer architectures is an important consideration when choosing a model for a specific application. Overall, both LSTM and Transformer are powerful neural network architectures that can be used for time series data regression tasks, and the choice between them may depend on the specific characteristics of the data and the task requirements.




\section{Conclusions and Future Outlook} \label{sectionconclusion}
This paper presents the use of a Transformer-based nonlinear compensation method for long-haul coherent optical communication system, which achieves a high Q factor gain as compared to DBP, FCNN and BiLSTM. On the basis of Q factor measurements, we have proved that the Transformer-based fiber nonlinear compensation model has excellent nonlinear fitting, accuracy, and generalizability achieving a Q factor gain of up to 1.94 dB wrt LE and 1.65 dB compared to the state-of-the-art DBP. Furthermore, the model has proven to be reliable in a wide range of transmission scenarios, adapting to different fiber lengths and launched optical powers. The flexibility and versatility of the Transformer architecture makes it a promising approach for future signal processing applications. These results showed that Transformer is a good architecture for NLC, which opens up its applications in the optical communications domain. Our future investigations will be to assess the Transformer model's adeptness in addressing the complexities posed by multi-channel nonlinear effects. Specifically, we will explore the model's potential to mitigate the intricate challenges presented by phenomena like cross-phase modulation and four-wave mixing. By rigorously evaluating the Transformer's capability to account for these inter-channel nonlinear interactions, we aspire to extend its applicability to scenarios characterized by diverse and intricate optical signal interactions. We also intend to delve into the utilization of more robust feature extractors, such as the restricted Boltzmann machine (RBM). By harnessing the capabilities of the RBM, we aim to create enhanced and discriminative features that can be seamlessly integrated as inputs to the Transformer model.


%

\section*{Acknowledgment}
The authors thank VPIphotonics and Optica foundation for providing training and access to VPI Photonics Design Suite.

\end{document}